\documentstyle[aps, preprint]{revtex}
\begin{document}
\draft

\newcommand{\bq}{\begin{equation}}
\newcommand{\eq}{\end{equation}}
\newcommand{\bqn}{\begin{eqnarray}}
\newcommand{\eqn}{\end{eqnarray}}
\newcommand{\nb}{\nonumber}
\newcommand{\lb}{\label}

\title{Gravitational Collapse of Perfect Fluid}
\author{J.F. Villas da Rocha${}^{a}$, Anzhong Wang${}^{b}$
 and N.O. Santos${}^{c}$ }

\address{ ${}^{a}$ Departmento de Astronomia Gal\'atica e 
Extra-Gal\'atica, Observat\'orio Nacional~--~CNPq, 
Rua General Jos\'e Cristino 77, S\~ao Crist\'ov\~ao, 20921-400 
Rio de Janeiro~--~RJ, Brazil\\
$^{b}$ Departamento de F\' {\i}sica Te\' orica,
Universidade do Estado do Rio de Janeiro,
Rua S\~ ao Francisco Xavier 524, Maracan\~ a,
20550-013 Rio de Janeiro~--~RJ, Brazil\\
${}^{c}$ Laboratorio de Astrofisica e Radioastronomia,
Centro Regional Sul de Pesquisas Espaciais - INPE/MCT,
LACESM, Cidade Universitaria,
97105-900 Santa Maria ~--~RS, Brazil}

\date{ November 10, 1998}

\maketitle

\begin{abstract}

The spherical gravitational collapse of a compact packet consisting of
perfect fluid is studied.  The spacetime outside the fluid packet 
is described by the out-going Vaidya radiation fluid. It is 
found that when the collapse has continuous
self-similarity  the formation of black holes always 
starts with zero mass, and when the collapse has no  
self-similarity, the formation of black
holes always starts with a finite non-zero mass.
The packet is usually accompanied by a thin matter shell. The effects
of the shell on the collapse are also studied.

\end{abstract}

\vspace{.4cm}


\section{Introduction}

Critical phenomena in gravitational collapse have attracted much 
attention \cite{Gu1997} since the pioneering work of Choptuik
\cite{Ch1993}. From the known results the following
emerges \cite{WO1997}: In general critical collapse of a
fixed matter field can be divided into three different classes
according to the self-similarities that the critical solution possesses. 
If the critical solution has no self-similarity, continuous or discrete,
the formation of black holes always starts with a  mass gap
(Type I collapse), otherwise it will start with zero mass 
(Type II collapse), and the mass of black holes takes
 the scaling form 
$
M_{BH} \propto (P - P^{*})^{\gamma}$, 
where $P$ characterizes the strength
of the initial data.
In the latter case, the collapse can be further
divided into two subclasses according to whether 
the critical solution has continuous self-similarity (CSS)  
or discrete self-similarity (DSS). Because of this difference, 
the exponent $\gamma$ is usually
also different. Whether the critical solution is CSS, DSS, or none of
them, depending on both the matter field
and the regions of the initial data space 
\cite{Gu1997}.  The co-existence of Type I and Type II collapse 
was first found in the SU(2) Einstein-Yang-Mills case \cite{CCB1996},
and later extended to both the Einstein-scalar case \cite{CLH1997}
and the Einstein-Skyme case \cite{BC1998},
while the co-existence of CSS and DSS critical solutions was 
 found in the Brans-Dicke theory \cite{LC1996}. The uniqueness of the
exponent in Type II collapse is well understood in terms of perturbations 
\cite{HKA1996}, and is closely related to the fact that the critical 
solution has only one unstable mode. This property now is considered as 
the main criterion for a solution to be critical \cite{Gu1997}.

While the uniqueness of the exponent $\gamma$ crucially 
depends on the numbers of the unstable
modes of the critical solution, that whether or not the formation
of black holes starts with a mass gap  
seemingly only depends on whether
the spacetime has self-similarity or not. Thus, even the
collapse is not critical, if a spacetime has CSS or DSS,
 the formation of black holes may still turn on
with zero mass. If this speculation is correct, it may have  
profound physical implications. For example, if Nature forbids
the formation of zero-mass black holes, which are essentially naked
singularities \cite{Gu1997}, it means that Nature forbids solutions
with self-similarity \cite{Si1996}. To study this problem
in its generality term, it is found difficult. In \cite{WRS1997},
which will be referred as Paper I, gravitational
collapse of massless scalar field and radiation fluid is studied, and
it was found that when solutions have CSS, the formation of black holes
indeed starts with zero-mass, while when solutions have no 
self-similarity it starts with a mass gap. 

In this Letter, we shall generalize the studies given in Paper I
to the case of perfect fluid with the equation of state $p = k
\rho$, where $\rho$ is the energy density of the fluid, $p$ the
pressure, and $k$ an arbitrary constant, subjected to $0 \le k \le 1$.
We shall show that the emerging results are consistent with the ones
obtained in Paper I. Specifically, we shall present two 
classes of exact solutions to the
Einstein field equations that represent spherical gravitational
collapse of perfect fluid, one has CSS, and the other has neither 
CSS nor DSS. It is found that such
formed black holes usually do not have finite masses. To remedy
this shortage, we shall cut the spacetime along a time-like
hypersurface, say, $r = r_{0}(t)$, and then join the internal region
$r \le r_{0}(t)$ with an asymptotically flat out-going Vaidya 
radiation fluid, using Israel's method \cite{Is1965}. It turns out that
in general such a junction is possible only when 
a thin matter shell is present
on the joining hypersurface \cite{BOS1989}. Thus, the finally 
resulting spacetimes
will represent the collapse of a compact packet of
perfect fluid plus a thin matter shell. The effects of the thin shell
on the collapse are also studied. It should be noted that
 by properly choosing the
solution in the region $r \ge r_{0}(t)$, in principle one 
can make the thin shell disappear, although in this Letter we shall not
consider such  possibilities. The notations will closely follow
the ones used in Paper I.

\section{Exact Solutions Representing Gravitational Collapse of
perfect Fluid}

In this section, we shall present two classes
 of solutions to the Einstein
field equations, 
$$
R_{\mu\nu} - \frac{1}{2}g_{\mu\nu}R =
 (\rho + p)u_{\mu}u_{\nu} - p g_{\mu\nu},
$$
where $u_{\mu}$ is the four-velocity of the perfect fluid considered. 
The general metric of spherically symmetric spacetimes that are conformally
flat is given by \cite{WRS1997}
\bq
\lb{eq2}
ds^{2} = G(t, r) \left[ dt^{2} - h^{2}(t, r)\;\left(dr^{2} 
+ r^{2}d\Omega^{2}\right)\right],
\eq
where $d\Omega^{2} \equiv d\theta^{2} + \sin^{2}\theta d\varphi^{2}$,
$\{x^{\mu}\}\equiv \{t, r, \theta, \varphi\}\; (\mu = 0, 1, 2, 3)$ are
the usual spherical coordinates, and 
\bq
\lb{eq3}
h(t, r) = \left\{
\begin{array}{c}
1, \\
{\left(f_{1}(t) + r^{2}\right)}^{-1},
\end{array}
\right.
\eq
with $f_{1}(t) \not= 0$. The Friedmann-Robertson-Walker (FRW) metric
corresponds to $G(t, r) = G(t)$ and $f_{1}(t) = Const.$ 
The corresponding Einstein field equations
are given by Eqs.(2.20) - (2.23) in Paper I. Integrating those equations,
we find two classes of solutions. In the following, we shall
present them separately.   

{\bf $\alpha$) Class A Solutions:} The first class of the solutions 
is given by 
\bqn
\lb{eq4}
G(t, r)  &=& (1 - Pt)^{2\xi}, \;\;\; \;\;\; \;\;\; h(r) = 1,\nb\\
 p = k \rho &=&  3 k \xi^2 P^2\left( 1 - Pt  
\right)^{-2(\xi+1)},\;\;
u_{\mu} = \left(1 - Pt  \right)^{-\xi} \delta^{t}_{\mu},
\eqn
where $P$ is a constant and characterizes the strength of the solutions
(See the discussions given below),
and $\xi \equiv 2/(1 + 3k)$. This class of solutions is actually the
FRW solutions and has CSS symmetry \cite{PZ1996}. However, in this Letter
we shall study them in the context of gravitational collapse.
  
To study the physical properties of these 
solutions, following Paper I
we consider the following physical quantities,
\bqn
\lb{eq9}
m^{f}(t, r) &\equiv& \frac{R}{2}(1 + 
R_{,\alpha}R_{,\beta} g^{\alpha \beta})
= \frac{\xi^{2}P^{2}r^{3}}{2(1 - Pt)^{2 - \xi}},\nb\\
{\cal{R}} &\equiv& R_{\alpha \beta\gamma\delta}
R^{\alpha \beta\gamma\delta} = \frac{18\xi^{2}(1+\xi^{2})P^{4}}
{(1 - Pt)^{4(1+\xi)}},
\eqn
where $R$ is the physical radius of the two sphere $t, r = Const.$, and
$m^{f}(t, r)$ is the local mass function \cite{PI1990}. From Eq.(\ref{eq9})
we can see that the spacetime is singular on the space-like hypersurface
$t = P^{-1}$. The nature of the singularity depends on the signature of the
parameter $P$. When $P < 0$, it is naked, and the corresponding solutions
represent white holes \cite{WRS1997}. When $P = 0$, the
singularity disappears and the corresponding spacetime is Minkowski.
When $P > 0$, the singularity hides behind the apparent horizon, which
locates on the hypersurface,
\bq
\lb{eq10}
r = r_{AH} \equiv \frac{1 - P t}{\xi P},
\eq
with $r_{AH}$ being a solution of the equation $R_{,\alpha}R_{,\beta}
g^{\alpha \beta} = 0$. Thus, in the latter case the solutions represent the
formation of black holes due to the gravitational collapse of the fluid.
The corresponding Penrose diagram is similar to that given by Fig.1(a)
in Paper I. Note that, although  the spacetime
singularity is always space-like, the nature of the apparent horizon
depends on the choice of the parameter $k$. In fact,  
 when $1/3 < k \le 1$, it is space-like; when $ k = 1/3$, it is
null; and when $ 0 \le k < 1/3$, it is time-like. 
Substituting Eq.(\ref{eq10}) into Eq.(\ref{eq9}), we find that $m^{f}_{AH}(t,
r_{AH}) = (P\xi)^{\xi}r_{AH}^{1 + \xi}$. Thus, as $r_{AH} \rightarrow +
\infty$, we have $m^{f}_{AH} \rightarrow + \infty$. That is, the total mass
of the black hole is infinitely large.  To get a physically reasonable
model, one way is to cut the spacetime along a
time-like hypersurface, say, $r = r_{0}(t)$, and then join the part $r \le
r_{0}(t)$ with one that is asymptotically flat \cite{WO1997}. 
We shall consider such  junctions  in the next section.

{\bf $\beta$) Class B Solutions:} The second class of solutions 
are given by
\bq
\lb{eq11}
G(t,r) = \sinh^{2\xi}\left[2\alpha{\xi^{-1}}
(t_{0} - \epsilon t)\right], \;\;\; 
h(r) = (r^{2} - \alpha^{2})^{-1},
\eq
where $\epsilon = \pm 1$, $\xi$ is defined as in Eq.(\ref{eq4}),
$t_{0}$ and $\alpha (\equiv \sqrt{-f_{1}})$ are 
constants. Introducing
a new radial coordinate $\bar{r}$ by $d\bar{r} = h(r)dr$,   
the corresponding metric can be written in the form
\bq
\lb{eq13}
ds^{2} = \sinh^{2\xi}[2\xi^{-1}(t_{0} - \epsilon t)]\left\{
dt^{2} - d{r}^{2} - \frac{\sinh^{2}(2
{r})}{4}
d^{2}\Omega\right\}.
\eq
Note that in writing the above equation we had, without
loss of generality, chosen $\alpha = 1$, and dropped the bar 
from $\bar{r}$. The energy density and
four-velocity of the fluid are given, respectively, by
\bq
\lb{eq14}
 p = k \rho = 12k
\sin^{-2(\xi+1)}[2(t_{0} - \epsilon t)/\xi],\;\;
u_{\mu} = \sin^{-\xi}[2(t_{0} - 
\epsilon t)/\xi]\delta^{t}_{\mu},
\eq
while the relevant physical quantities are given by 
\bqn
\label{eq15}
m^{f}(r,t) &=&   {1 \over 4 } \sinh^3(2  {r} )
\sinh^{\xi -2} {\left[ {2\xi^{-1}} {( t_0 - \epsilon t)} 
\right]}, \nb\\
{\cal{R}}  &=&  {288\left(1+ \xi^2 \right) \xi^{-2}} 
\sinh^{-4 (\xi +1)}  
{\left[ {2 \xi^{-1}} {\left( t_0 - \epsilon t\right)} 
\right]}.
\eqn
The apparent horizon now is located at
\bq
\lb{eq16}
r = r_{AH} \equiv { \xi^{-1} } (t_0 - \epsilon t).
\eq
From Eq.(\ref{eq15}) we can see that the solutions are singular on the
hypersurface $t = \epsilon t_{0}$. When $\epsilon = - 1$ it can be
shown that the corresponding solutions represent cosmological models
with a naked singularity at the initial time $t = - t_{0}$, while when
$\epsilon = + 1$ the singularity is hidden behind the apparent horizon
given by Eq.(\ref{eq16}), and the solutions represent  
the formation of black holes due to the collapse of the fluid.
In the latter case the total mass of black holes is also infinitely large.
To remedy this shortage, in the next section we
shall make ``surgery" to this spacetime, so that
 the resulting black holes have finite masses.

\section{Matching the Solutions with Outgoing Vaidya Solution}

In order to have the black hole mass finite, we shall first cut the 
spacetimes represented by the solutions given by Eqs.(\ref{eq4}), and
(\ref{eq13}) along a time-like hypersurface, and then join the internal
part with the out-going Vaidya radiation fluid.  
In the present two cases since the perfect fluid is
comoving, the hypersurface can be chosen as $r = r_{0} = Const.$ 
Thus, the metric in the whole spacetime can be written in the 
form
\bq 
\label{eq17}
ds^2  = \left\{
\begin{array}{c}
A(t,r)^2 dt^2 - B(t,r)^2dr^{2} -  
C(t, r)^2d\Omega^2, \;(r \le r_{0}), \\
\left( 1- \frac{2m(v)}{R}\right)dv^2 + 2dvdR^2 
- R^2d\Omega^2, \; (r \ge r_{0}),
\end{array}
\right.
\eq
where the functions $A(t, r),\; B(t, r)$ and $C(t, r)$ can be
read off from Eqs.(\ref{eq2}), (\ref{eq4}) and (\ref{eq13}). On  
 the hypersurface $r= r_{0}$ the metric reduces to 
\bq 
\label{eq18}
ds^2 \left|_{r = r_{0}} = \right. g_{ab}d\xi^{a}d\xi^{b} =
d\tau^2  -  R(\tau)^2d\Omega^2,
\eq
where $\xi^{a} = \{\tau, \theta, \varphi\}$ are the intrinsic
coordinates of the surface, and $\tau$ is defined by 
\bq
\lb{eq19}
d\tau^{2} = A^{2}(t, r_{0})dt^{2} 
= \left(1 - \frac{2M(\tau)}{R}\right)dv^{2}
+ 2dvdR,
\eq
where $v$ and $R$ are functions of $\tau$ on the surface, 
and $R(\tau) \equiv C(t, r_{0}),\; M(\tau) \equiv m(v(\tau))$.  
The extrinsic curvature oon the two sides
of the surface  defined by
\bq
\lb{eq20}
K^{\pm}_{ab} = - n^{\pm}_{\alpha}\left[
\frac{\partial^{2}x^{\alpha}_{\pm}}{\partial \xi^{a} 
\partial \xi^{b}}
- \Gamma^{\pm \alpha}_{\beta\delta}\frac{\partial 
x^{\beta}_{\pm}}
{\partial \xi^{a}}\frac{\partial x^{\delta}_{\pm}}
{\partial \xi^{b}}\right],
\eq
has the following non-vanishing components \cite{Ch1997}
\bqn
\lb{eq21}
K_{\tau\tau}^{-} &=& - \frac{{\dot t}^2 A_{,r} A}{B},\;\;
K_{\theta\theta}^{-} = \sin^{- 2}\theta K_{\varphi\varphi}^{-}=
\frac{C_{,r} C}{B},\nb\\ 
K_{\tau\tau}^+ &=& \frac{\ddot v}{\dot v} -
 \frac{{\dot v} M(\tau)}{R^2},\;\;
K_{\theta\theta}^+ = \sin^{-2}\theta K_{\varphi\varphi}^+
= R\left\{ 
{\dot v}\left( 1- \frac{2M(\tau)}{R}\right)  + \dot{R} \right\},
\eqn
where  $\dot{t} \equiv dt/d\tau,\; (\;)_{,\mu} \equiv
\partial(\;)/\partial x^{\mu}$ and  $n_{\alpha}^{\pm}$ are the normal
vectors defined in the two faces of the surface. Using the expression 
\cite{Is1965}
\bq
\lb{eq22}
\left[K_{ab}\right]^{-} - g_{ab}\left[K\right]^{-} = - 8\pi \tau_{ab}
\eq
we can calculate the surface energy-momentum tensor $\tau_{ab}$, where
$\left[K_{ab}\right]^{-} = K_{ab}^{+} - K_{ab}^{-},\; [K]^{-} = g^{ab}
\left[K_{ab}\right]^{-}$, and $g_{ab}$ can be read off from 
Eq.(\ref{eq18}). Inserting Eq.(\ref{eq21}) 
into the above equation, we find that $\tau_{ab}$ 
 can be written in the form
\bq
\lb{25}
\tau_{ab} = \sigma w_a w_b  + 
\eta \left(\theta_a \theta_b + \phi_a \phi_b\right),
\eq
where $w_{a}, \; \theta_{a}$ and $\phi_{a}$ are unit vectors
defined on the surface, given respectively by
$
w_a = \delta^{\tau}_a,\;
\theta_a = R\delta^\theta_a,\;
\phi_a = R \sin\theta\delta^\varphi_a$,
and $\sigma$ can be interpreted  as the surface energy density, 
$\eta$ the tangential pressure, provided that they satisfy 
certain energy conditions \cite{HE1973}. In the present case $\sigma$
and $\eta$ are given  by  
\bqn
\lb{eq27}
\sigma  &=&  \frac{1}{4\pi R}
\left\{\dot{R} - \frac{1}{\dot{v}} + J'(r_{0})\right\},
  \nb\\
\eta  & = & \frac{1}{16\pi R \dot{v}}
\left\{  \dot{v}^{2} - 2\ddot{v}R - 2\dot{v}J'(r_{0}) + 1
  \right\},
\eqn
where $J(r) = r$ for Class A solutions, and $J(r) = \sinh(2r)/2$ for
Class B solutions, and a prime denotes the ordinary differentiation
with respect to the indicated argument. Note that 
in writing Eq.(\ref{eq27}) we had used Eq.(\ref{eq19}), from which  
it is found that the total mass of the collapsing ball, 
which includes the contribution 
from both the fluid and the shell, is given by 
\bq
\lb{eq28}
M(\tau) = \frac{R}{2\dot{v}^{2}}
\left(\dot{v}^{2} + 2\dot{v}\dot R - 1\right).
\eq  
To fix the spacetime outside
the shell we need to give the equation of state of the shell. In 
order to minimize the effects of the shell on the collapse, in the
following we shall consider the case $\eta = 0$, which reads
\bq
\lb{eq29}
{\dot v}^2 - 2 {\ddot v } R - 2 J'(r_{0}){\dot v}  + 1   =  0.
\eq
To solve the above equation, let us consider the two classes
of solutions separately. 

\subsection{Class A Solutions}

In this case, it can be shown that Eq.(\ref{eq29}) has the first
integral,
\bq
\lb{eq30}
\dot{v}(\tau) =  \frac{x - 2(v_{0} - 1)R_{0}}{x - 2v_{0}R_{0}},
\eq
where 
$R(\tau) \equiv R_{0}x^{\xi},\;
R_{0} \equiv r_{0}P^{\frac{\xi}{\xi + 1}},\;
x \equiv \left[(\xi + 1)(\tau_{0} - \tau)\right]^{\frac{1}{\xi + 1}}
$, and $v_{0}$ and $\tau_{0}$ are integration constants. 
Substituting the above expressions into Eq.(\ref{eq28}), we find that 
\bq
\lb{eq31}
M(x) = \frac{R_{0}^{2}x^{\xi -1}}{[x  -2(v_{0} - 1)R_{0}]^{2}}
 \left\{(2 - \xi)x^{2} + 2(\xi - 1)(2v_{0} - 1)R_{0}x 
 + 4 \xi v_{0}(1 - v_{0})R_{0}^{2}\right\}.
\eq
At the moment $\tau = \tau_{AH}\; $(or $x = x_{AH} = \xi R_{0}$), 
the shell collapses inside the apparent horizon. Consequently, the 
total mass of the formed black hole is given by
\bq
\lb{eq32}
M_{BH} \equiv M(x_{AH}) =
\frac{\xi^{\xi}r_{0}^{1+\xi}P^{\xi}}
{[\xi - 2(v_{0} - 1)]^{2}}\left\{\xi(2-\xi) + 2(\xi -1)(2v_{0} - 1)
  + 4 v_{0}(1 - v_{0})\right\},
\eq
which is finite and can be positive
by properly choosing the parameter $v_{0}$ for any given $\xi$.
The contribution of the fluid and the thin shell to the black 
hole mass is given, respectively, by \footnote{
While a unique definition of the total mass of a 
thin shell is still absent,
here we simply define it as $m^{shell}_{BH} \equiv 4 \pi R^{2} \sigma$. 
Certainly we can
equally use other definitions, such as $m^{shell}_{BH} 
 \equiv M_{BH} - m^{f}_{BH} $,
but our final conclusions will not depend on them.
}
\bqn
\lb{eq33}
m^{f}_{BH} &\equiv& m^{f}_{AH}(\tau_{AH}) = 
\frac{\xi^{\xi}r_{0}^{1+\xi}}{2}P^{\xi},\nb\\
m^{shell}_{BH} &\equiv& 4\pi R^{2}(\tau_{AH})\sigma(\tau_{AH})
= \frac{\xi^{\xi}r_{0}^{1+\xi}(2v_{0} - \xi)}
{\xi - 2(v_{0} - 1)}P^{\xi}.
\eqn
From the above equations we can see that all the three masses are 
proportional to $P$, the parameter that characterizes the strength
of the initial data of
 the collapsing ball. Thus, when the initial data is very weak
($P \rightarrow 0$), the mass of the formed black hole is very small
($M_{BH} \rightarrow0$). In principle, by properly tuning the parameter
$P$ we can make it as small as wanted. Recall that now the solutions 
have CSS. It should be noted that
due the the gravitational interaction between the collapsing fluid 
and the thin shell, we have $M_{BH} \not= m^{f}_{BH} + m^{shell}_{BH}$, 
unless $\xi = 2$, which corresponds to null dust. In the latter case,
it can be shown that by choosing $v_{0} = 1$ we can make the thin shell
disappear, and the collapse is purely due to the null fluid. Like the cases
with thin shell, by properly tuning the parameter $P$ we can make black holes
with infinitesimal mass. When $\xi = 1/2$ or $1$, which corresponds,
respectively, to the massless scalar field or to radiation fluid, the
solutions reduce to the ones considered in \cite{WRS1997}. 

Note that although the mass of black holes takes
 a scaling form in terms of $P$, the exponent $\gamma$
is not uniquely defined. This is because 
in the present case the solution with $P = P^* = 0$  
separates black holes from white holes,
and the latter is not the result
of gravitational collapse. Thus,  the solutions considered here
do not really represent the critical collapse. As a result,  
we can replace $P$ by any function $P(\bar P)$, and for each 
of such replacements, we will have a different $\gamma$ \cite{Gundlach1996}.
However, such replacements do not change the fact that by properly
tuning the parameter we can make black holes with masses as small
as wanted.

\subsection{Class B Solutions}

In this case, the first integral of Eq.(\ref{eq29}) yields
\bq
\lb{eq34}
\dot{v} = \cosh(2r_{0}) - \sinh(2r_{0})\tanh(t + t_{1}),
\eq
where $t_{1}$ is an integration constant. At the moment $ t = t_{AH}$
the whole ball collapses inside the apparent horizon, and the 
contribution of the fluid and the shell to the total mass of the just 
formed black holes are given, respectively, by
\bqn
\lb{eq35}
m^{f}_{BH} &\equiv& m^{f}_{AH}(\tau_{AH}) = 
\frac{1}{4}\sinh^{\xi +1}(2r_{0}),\nb\\
m^{shell}_{BH} &\equiv& 4\pi R^{2}(\tau_{AH})\sigma(\tau_{AH})
= - \frac{1}{2}\sinh^{\xi+1}(2r_0)
\frac{\cosh[t_1-t_0+\xi r_0]}{\cosh[t_1-t_0- (2-\xi)r_0]}
\eqn
From the above expressions we can see that for any given $r_{0},\;
m^{f}_{BH}$ and $ m^{shell}_{BH}$ are always finite and non-zero. Thus,
in the present case black holes start to form with a mass gap. 
It should be noted that although $m^{f}_{BH}$ is positive, $m^{shell}_{BH}$
is negative.  
This undesirable feature makes the model very unphysical. One may look
for other possible junctions. However, since the fluid is co-moving,
one can always make such junctions across the hypersurface 
$r = r_{0} = Const.$. Then, the contribution of the collapsing 
fluid to the total mass of black holes will be still given by 
Eq.(\ref{eq35}), and the total mass of the formed balck hole
then can be written in the form,
$$
M_{BH} = \frac{1}{4}\sinh^{\xi +2}(2r_{0}) + M_{BH}^{rest},
$$
where $M_{BH}^{rest}$ denotes the mass contribution 
of the rest part of the spacetime, which is non-negative for 
any physically reasonable junction. Therefore, in this case for any
physical junction, the total mass of black holes will be finite and
non-zero.

\section{CONCLUSION} 

In this Letter, we have studied two classes of solutions to the Einstein
field equations, which represent the spherical gravitational collapse of
a packet of perfect fluid, accompanied usually by a thin matter shell. 
The first class of solutions has CSS, and black holes
always start to form with a zero-mass, while the second class has 
neither CSS nor DSS, and the formation of black holes always starts with a
mass gap. The existence of the matter shell does not affect our
main conclusions.  These solutions provide further evidences to support our
speculation that {\em the formation of black holes always starts 
with zero-mass for the collapse with self-similarities, CSS or DSS}. 
 
It should be noted that none of these two classes of solutions given 
above represent critical collapse. Thus, whether the formation
of black holes starts with zero mass or not is closely related to the
symmetries of the collapse (CSS or DSS), rather than to whether the
collapse is critical or not.

\section*{Acknowledgment}

The financial assistance from CAPES (JFVR), CNPq (AW, NOS), and FAPERJ (AW)
is gratefully acknowledged.


\begin{thebibliography}{99}


\bibitem{Gu1997} C. Gundlach,``{\em Critical phenomena in 
gravitational collapse},'' $gr-qc/9712084$, preprint (1997); 
M.W. Choptuik, ``{\em The (unstable) threshold of black hole formation},"  
$gr-qc/9803075$, preprint (1998).

\bibitem{Ch1993} M.W. Choptuik, Phys. Rev. Lett. {\bf 70}, 9 (1993).

\bibitem{WO1997} A.Z. Wang and H.P. de Oliveira, Phys. Rev. 
{\bf D56}, 753 (1997).

\bibitem{CCB1996}  M.W.  Choptuik, T. Chmaj, and P. Bizo\'n, Phys. Rev. 
Lett. {\bf 77}, 424 (1996); C. Gundlach, Phys. Rev. {\bf D55}, 6002 (1997).

\bibitem{CLH1997} M.W. Choptuik, S.L. Liebling, and E.W. Hirschmann,
Phys. Rev. {\bf D55}, 6014 (1997); P. Brady, C.M. Chambers, and 
S.M.C.V. Goncalves, {\em ibid.} {\bf D56}, 6057 (1997).

\bibitem{BC1998} P. Bizo\'n and T. Chmaj, ``{\em Critical collapse of
 Skymions},"  $gr-qc/9801012$, preprint (1998).
 
\bibitem{LC1996} S.L. Liebling and M.W.  Choptuik, Phys. Rev. 
Lett. {\bf 77}, 1424 (1996); E.W. Hirschmann and D.M.  Eardley,
Phys. Rev. {\bf D56}, 4696 (1997).

\bibitem{HKA1996} T. Hara, T. Koike, and S. Adachi, ``{\em Renormalization 
group and critical behavior in gravitational collapse},"  
$gr-qc/9607010$, preprint (1996). 

\bibitem{Si1996} T.P. Singh, ``{\em Gravitational Collapse and Cosmic
Censorship}," $gr-qc/9606016$, preprint (1996).

\bibitem{WRS1997} A.Z. Wang, J.F.Villas da Rocha, and N.O. Santos, 
 Phys. Rev. {\bf D56}, 7692 (1997).

\bibitem{Is1965} W. Israel, Nuovo Cimento, {\bf B44}, 1 (1966); {\em
ibid.}, {\bf B48}, 463(E) (1967).

\bibitem{BOS1989} W. Israel, Nuovo Cimento, {\bf B44}, 1 (1966); {\em
ibid.}, {\bf B48}, 463(E) (1967).

\bibitem{PZ1996} D. Pollney and T. Zannias, Ann. Phys. {\bf 252}, 241 (1996).

\bibitem{PI1990}E. Poisson and W. Israel,  Phys. Rev. {\bf D41}, 1796 (1990).

\bibitem{Ch1997} R. Chen, Mon. Not. R. Astron. Soc. {\bf 288}, 589 (1997).

\bibitem{Vd1951}  P.C. Vaidya, Proc. Indian Acad. Sc., {\bf A33}, 264 (1951). 

\bibitem{HE1973} S.W. Hawking and G.F.R. Ellis, {\em The Large Scale
Structure of Spacetime}, (Cambridge University Press, Cambridge,
1973).
 
\bibitem{Gundlach1996} C. Gundlach,``{\em Critical Phenomena in Gravitational
collapse}'' $gr-qc/9606023$, preprint (1996).

\end {thebibliography}
\end{document}